\documentclass[aps, nofootinbib,longbibliography,superscriptaddress, preprint]{revtex4-1}

\usepackage{fancyhdr, amssymb, cancel, amsmath, graphicx, pgfplots, tikz}

\newcommand{\cQ}{\mathcal{Q}}
\usepackage[colorlinks=true, urlcolor=violet, linkcolor=blue, citecolor=red, hyperindex=true, linktocpage=true]{hyperref}

\newcommand{\comment}[1]{{\bf\color{blue}[#1]}}

\begin{document}
\title{Scale-invariant hyperscaling-violating holographic theories and the resistivity of strange metals with random-field disorder}
\author{Andrew Lucas}
    \email{lucas@fas.harvard.edu}
\affiliation{Department of Physics, Harvard University, Cambridge MA
  02138, USA} 

\author{Subir Sachdev}
  \email{sachdev@g.harvard.edu}
\affiliation{Department of Physics, Harvard University, Cambridge MA
  02138, USA} 

\author{Koenraad Schalm}
  \email{kschalm@lorentz.leidenuniv.nl}
\affiliation{Department of Physics, Harvard University, Cambridge MA
  02138, USA} 
\affiliation{Institute Lorentz, Leiden University, PO
  Box 9506, Leiden 2300 RA, The Netherlands}

\begin{abstract}

We compute the direct current resistivity of a scale-invariant, $d$-dimensional strange metal with dynamic critical exponent $z$ and 
hyperscaling-violating exponent $\theta$, weakly perturbed by a scalar operator coupled to random-field disorder that locally breaks a $\mathbb{Z}_2$ symmetry.   Independent calculations via Einstein-Maxwell-Dilaton holography and memory matrix methods lead to the same results.   We show that random field disorder has a strong effect on resistivity, and leads to a short relaxation time for the total momentum.
In the course of our holographic calculation we use a non-trivial dilaton coupling to the disordered scalar,  allowing us to study a strongly-coupled scale invariant theory with $\theta \ne 0$.   Using holography, we are also able to determine the disorder strength at which perturbation theory breaks down.   Curiously, for locally critical theories this breakdown occurs when the 
resistivity is proportional to the entropy density, up to a possible logarithmic correction.  

\end{abstract}

\maketitle
\tableofcontents

\section{Introduction}
One of the remarkable puzzles in quantum critical phases is the universality of the resistivity across widely different systems. In particular strange metals exhibit almost exclusively a dc-resistivity that scales linear in temperature. This is in contrast to the array of models that exist for quantum critical systems. A wide class of such quantum critical models 
can be characterized by
non-trivial dynamic critical (``Lifshitz'') exponent $z$ associated with the relative scaling of time and space $(t\sim x^z$), and a hyperscaling-violating exponent $\theta$ corresponding to the deviation of the scaling of the low-energy critical degrees of freedom from pure dimensional arguments: {\em i.e.\/} the degrees of freedom ``effectively live in'' (spatial) dimension $d-\theta$ \cite{sachdevbook}.\footnote{For example, in a theory with a Fermi surface, with low energy excitations described by chiral fermions, $\theta=d-1$.} 
%
%

At the same time the intimate tie-in of the dc-resistivity with translational symmetry breaking allows for a universal mechanism to emerge {\em if} there is a dominant such mechanism at low energies. One such mechanism is random-field disorder acting on a scalar order parameter, which has recently been shown to relax momentum much more rapidly than disorder coupled to charged fermionic excitations near the Fermi surface \cite{raghu}. In this paper, we will study its effects on the resistivity in a hyperscaling violating Lifshitz quantum critical system. The difficulty is that most such theories are thought to be (strongly) interacting in the regime of interest. We resort to two well-established techniques that can address the charge dynamics nevertheless: the memory matrix method \cite{forster} and gauge-gravity duality \cite{hartnoll, mcgreevy, sachdev}, with generalizations to Lifshitz \cite{kachru} and hyperscaling-violating \cite{gubserrocha,charmousis,trivedi,lsb,harrison} geometries.  Despite the fact that most systems of
interest have explicit Fermi surfaces, and there are no
explicit Fermi surfaces in the holographic computation,
corrections due to fermion scattering near the Fermi surface
are subleading \cite{raghu}. The latter gives us an explicit description of a strongly coupled hyperscaling violating Lifshitz quantum critical system in terms of a dual Einstein-Maxwell-Dilaton (EMD) system. Strictly put this gravity-dual only describes the large $N$-matrix limit of the quantum critical theory, but arguably the scaling behavior we are interested does not depend strongly on this. This is confirmed by the memory matrix computation, which works universally when translational symmetry is only weakly broken.\footnote{In principle the memory matrix method always works for a clean separation of fast and slow modes. In practice one needs to know the correlation functions of the slow modes, which are not always universal. If translational symmetry is weakly broken, however, then the universality of the energy-momentum current as a slow mode allows one to obtain universal analytic answers.}  
With a necessary refinement of EMD holography that we explain below, we show that these two approaches agree.
Our holographic computation shows that
  for random fields of typical size $\varepsilon$, which couple to a random field of dimension $\Delta$, the leading-order perturbative contribution to the d.c. resistivity is given by 
\begin{equation}
{\rho_{\mathrm{dc}} \sim \varepsilon^2 T^{2(1+\Delta-z)/z} + \mathrm{O}\left(\varepsilon^4\right). } \label{rhodc}
\end{equation}
Interestingly, this result is independent of the hyperscaling-violation exponent $\theta$.   Some limiting cases of this result have been obtained earlier \cite{hkms, hartnollimpure} by memory matrix methods.

From Eq. (\ref{rhodc}), we conclude that random-field disorder has an extremely strong effect on the resistivity in hyperscaling violating Lifshitz quantum critical systems.    This is in contrast to the recent result \cite{dsz} which studied the same question in the locally critical 
limit $z \rightarrow \infty$ of some holographic models and found the enticing identity $\rho_{\mathrm{dc}}\sim s$ (where $s$ is the thermal entropy density).    This scaling suggests a possible universal explanation for the linear-in-temperature resistivity of the strange metals.   However, we found that for finite Lifshitz scaling $z \neq \infty$ this identity does not hold.   Even at $z=\infty$, we recover this result in a rather curious way --- although this identity does not appear to follow from Eq. (\ref{rhodc}), we will see that $\rho_{\mathrm{dc}}\sim s$ precisely at the onset of the regime where disorder must be treated non-perturbatively.

Our refinement of EMD theory is to include non-trivial dilaton coupling into the action of the disordered scalar.   From the perspective of supergravity truncations, this is a natural coupling to include.   
This non-trivial dilaton coupling
allows us to construct a strongly coupled hyperscaling-violating theory via holography which {maintains scale invariant correlation functions}, a simple but important result which has been noted in \cite{hartong, gouteraux}.  We expand on this result by explicitly deriving correlation functions, as well as criteria for the relevance of operators in terms of their boundary dimension $\Delta$, both with and without disorder, in a hyperscaling-violating background.


\section{Scale-Invariant Hyperscaling-Violating Holography}
The Einstein-Maxwell-Dilaton models that can capture through gauge-gravity duality the physics of hyperscaling-violating quantum critical field theories are described by the action
\begin{align}
  \label{eq:1}
S_{\mathrm{EMD}} = \int \mathrm{d}^{d+2}x\; \sqrt{-g} \left(\frac{R-2(\partial\Phi)^2 - V(\Phi)}{2\kappa^2} - \frac{Z(\Phi)}{4e^2}F^2\right)
\end{align}
The deep infrared (IR) of these theories is controlled by the leading exponent of the arbitrary functions $V(\Phi)=-V_0 \mathrm{e}^{-\beta\Phi}+\ldots$ and $Z(\Phi)=Z_0 \mathrm{e}^{\alpha\Phi}+\ldots$. Truncating these functions to this exponent, the theory has black brane solutions dual to the hyperscaling violating quantum critical ground states 
supported by a 
charge density $\mathcal{Q}$ \cite{charmousis,trivedi,lsb}. 
These solutions have a metric  
\begin{equation}
\label{eq1}
\mathrm{d}s^2  = \frac{L^2}{r^2}\left[\frac{G(r)}{f(r)}\mathrm{d}r^2 - f(r)H(r) \mathrm{d}t^2 + \mathrm{d}\mathbf{x}^2\right].
\end{equation}with non-vanishing Maxwell flux  \begin{equation}
F = \frac{eL}{\kappa} h^\prime(r) \mathrm{d}r\wedge\mathrm{d}t
\end{equation}which sources a constant charge density $\mathcal{Q}$, which can be determined from Gauss' Law:\begin{equation}
\frac{e\kappa}{L^{d-1}} \cQ \equiv \hat{\cQ}=- Z\sqrt{-g}g^{tt}g^{rr} h^\prime, 
\end{equation} and a running dilaton \begin{equation}
\Phi = \frac{2}{\alpha} \left(d+\frac{\theta}{d-\theta}\right) \log \frac{r}{r_0}.
\end{equation}The functions $G$, $H$, $h^\prime$ scale with $r$ as follows:
\begin{subequations}
\begin{align}
G(r)&=G_0 r^{2\theta/(d-\theta)} \\
H(r)&= H_0 r^{-2d(z-1)/(d-\theta)}\\
h^\prime (r)&= h_0 r^{-1-d-dz/(d-\theta)}
\end{align}
\end{subequations}
The choice of the coefficients $\alpha,\beta$ in $Z(\Phi),V(\Phi)$ determine the dynamical critical exponent $z$ and hyperscaling violation as
\begin{subequations}\begin{align}
  \label{eq:19}
  \theta &= \frac{d^2\beta}{\alpha+(d-1)\beta} , \\
  z &=
  1+\frac{\theta}{d}+ \frac{8(d(d-\theta)+\theta)^2}{d^2(d-\theta)\alpha^2}
\end{align}\end{subequations}
The emblackening factor 
\begin{equation}
f(r)=  1-\left(\frac{r}{r_{\mathrm{h}}}\right)^{d(1+z/(d-\theta))}
\end{equation}
 places the system at a small but finite temperature $T$, related to the horizon radius $r_{\mathrm{h}}$ as 
\begin{equation}
r_{\mathrm{h}} \sim T^{-(1-\theta/d)/z} \mathcal{Q}^{-1/d}.
\end{equation}
The entropy density of this black hole manifestly exhibits hyperscaling violation:
\begin{equation}
s \sim r_{\mathrm{h}}^{-d} \sim T^{(d-\theta)/z}.
\end{equation}
In these coordinates $r\rightarrow r_{\mathrm{h}}$ captures the low energy regime of the dual QFT. At the opposite high energy end $r\rightarrow 0$, this IR solution can be connected to a complete asymptotically $\mathrm{AdS}_{d+2}$ EMD solution; see, {\em e.g.\/}, \cite{gubserrocha}.   We will not do so explicitly here.  Based on the insight that the radial direction corresponds to the Wilsonian scale of the theory, we shall cut-off the metric beyond this IR region and apply the holographic dictionary at this cut-off. Recalling that the IR geometry is completely controlled by the charge density $\cQ$: it sets the ultraviolet (UV) cut-off.
 As we will show later, $\rho_{\mathrm{dc}}$ is \emph{quantitatively} controlled by the IR geometry, and therefore a precise characterization of a UV completion is not necessary to understand scaling.  For more on matching procedures to asymptotically AdS spaces (in the UV), see \cite{faulkner}. 
For completeness let us mention that the ratio $L^d/\kappa^2$ roughly counts the degrees of freedom in the holographic theory, and must be large for classical gravity to be valid; $e$ is the unit of charge.

Via the holographic dictionary, additional fields in the bulk correspond to additional operators in the boundary theory.   For simplicity, we focus on 
bulk scalar fields.   
To quadratic order the action of an additional 
bulk scalar field $\psi$ will be of the form \begin{equation}
S_{\psi} = -\int \mathrm{d}^{d+2}x\; \sqrt{-g} \left(\frac{1}{2}(\partial \psi)^2 + \frac{B(\Phi)}{2}\psi^2\right).
\end{equation}
The function $B(\Phi)$ will be fine-tuned such that the correlation functions of $\psi$ exhibit manifest scaling behavior. 

To compute the Green's functions of the operator $\mathcal{O}$ dual to $\psi$, we solve the equation of motion for the bulk field $\psi$ in the background of Eq. (\ref{eq1}): \begin{equation}
\partial_M \left(\sqrt{-g}g^{MN}\partial_N\psi\right) = \sqrt{-g} B(\Phi)\psi.   \label{eqpsi0}
\end{equation}  To do this, we look for a simple choice of $B(\Phi)$.  In \cite{harrison}, the choice $B(\Phi)=m^2$ was used, and the result was a non-scale invariant quantum field theory, with a length scale set by the AdS radius $L$.    We make a different choice: it is easy to see that choosing \begin{equation}
B(\Phi(r)) \equiv  \frac{B_0}{L^2 G(r)} = \frac{B_0}{L^2 G_0} r^{-2\theta/(d-\theta)}~,
\end{equation}
equivalent to the choice $B(\Phi) \sim \mathrm{e}^{\gamma\Phi}$ with
\begin{equation}
\gamma = -\frac{\alpha \theta }{d(d-\theta) + \theta} =-\beta
\end{equation}
leads to a field theory which has both hyperscaling violation and scale invariance.   Generating scale invariance by adding dilaton couplings has been noted in \cite{hartong, gouteraux}.   Below we elaborate on the consequences.   At $T=0$, the zero-frequency 
solutions to the scalar equation of motion are now the usual Bessel functions
\begin{align}
\psi(k,r) = r^{\frac{1}{2}(d+\frac{dz}{(d-\theta)})}&\left(\alpha \mathrm{K}_{\frac{(d-\theta)}{2d}(\nu_+-\nu_-)}\left(C_{d,\theta}|k|r^{d/(d-\theta)}\right) \right.
\left.+
\beta \mathrm{I}_{\frac{(d-\theta)}{2d}(\nu_+-\nu_-)}\left(C_{d,\theta}|k|r^{d/(d-\theta)}\right) \right)
 \label{eq9}
\end{align} 
with $C_{d,\theta}=\frac{(d-\theta)\sqrt{G_0}}{d}$.  $\nu_-<\nu_+$ correspond to the power laws of the two solutions to the equations of motion at zero momentum and frequency: $\psi(\mathbf{k}=\mathbf{0},\omega=0,r) \sim r^{\nu_\pm}$ with \begin{equation}
2\nu_\pm \equiv  d+\frac{dz}{d-\theta} \pm \sqrt{\left(d+\frac{dz}{d-\theta}\right)^2 + 4B_0}.
\end{equation}
Following the usual dictionary of gauge-gravity duality, the ratio of the subleading solution ($\alpha=0$) in Eq. (\ref{eq9}) to the leading solution ($\beta=0$) in the limit $r \rightarrow 0$ gives the scaling behavior of the zero-frequency Green's functions 
of the operator $\mathcal{O}$ in the dual field theory.\footnote{We ignore subtleties between Euclidean and Lorentzian signature. For the scaling argument, this does not matter.} By construction the choice of $B(\Phi)$ gives the scaling solution: 
\begin{equation}
G(k,\omega=0) \sim k^{(1-\theta/d)(\nu_+-\nu_-)}.
\end{equation}
We denote by $\Delta$ the scaling dimension of the $\mathcal{O}$ operator. Then, 
in position space $\mbox{$G(x,t=0)$}\sim x^{-2\Delta}$, and we find
\begin{subequations}\begin{align}
\frac{d-\theta}{d}\nu_+ &= \Delta-\frac{\theta}{2}, \\
\frac{d-\theta}{d}\nu_- &= d+z-\Delta - \frac{\theta}{2}.
\end{align}\end{subequations}
The corresponding value of $B_0$ for any $\Delta$ can be straightforwardly found.   The requirement that a operator not be described by ``alternate quantization" (i.e. the requirement that $\nu_+>\nu_-$) is $\Delta > (d+z)/2$.
The condition that an insertion of the operator $\mathcal{O}$ in the boundary theory is a relevant perturbation, 
{\em i.e.\/} the scaling dimension of the {\em uniform\/} field $h_0$ is positive (where the insertion is $h_0 \int \mathrm{d}^{d+1} x \, \mathcal{O} (x)$)
is the same as the requirement that $\nu_->0$, which corresponds to 
\begin{equation}
\Delta<d+z - \frac{\theta}{2}.
\end{equation}  
We do not allow such uniform field insertions in the present paper.

\section{Conductivity with Random Field Disorder.}
We now discuss the impact of random field disorder on the
resistivity at zero frequency and momentum, $\rho_{\mathrm{dc}}$, in the field theory dual to the EMD black brane at finite $T$ and $\mathcal{Q}$, in two spatial dimensions. 
In a translation invariant background, the symmetry enforces 
 that $\rho_{\mathrm{dc}}=0$ \cite{forster}.  However, no realistic condensed matter system has true translational invariance.  One source of translational symmetry breaking is an underlying lattice, or any other periodic potential, whose effects on transport coefficients have been intensively studied recently with holography \cite{hartnollhofman,santos1,Liu:2012tr,hutasoit, santos2, santos3, lucas,ling, donos,Andrade:2013gsa,koushik,Donos:2014uba}.  
The other noted
 source of translational symmetry breaking is 
 disorder \cite{hkms,hartnollimpure,oai:arXiv.org:1102.2892,oai:arXiv.org:1201.6366,arean}. 
 Because disorder preserves translation symmetry on average, it is likely a much more tractable approach analytically. 
Indeed there are arguments that holographically the phenomology of disorder can be simply captured by a theory with massive gravity \cite{vegh, davison}, even non-perturbatively.\footnote{However, it may be the case that non-perturbative disorder causes horizon fragmentation, which certainly is not captured by massive gravity.   It is known that this is possible in $d=1$ \cite{brill}.   This is an important open question in higher dimensions.  We thank Sho Yaida for bringing up this possibility to us.} 

Below we will  
consider the limit of weak random-field disorder explicitly and compute the leading order temperature scaling of $\rho_{\mathrm{dc}}$ with two independent calculations:  first, using EMD holography, and second using memory matrix methods.
In the holographic calculation, we will  
exploit recent weak-field results \cite{blaketong1, blaketong2, gouteraux} to compute $\rho_{\mathrm{dc}}$, though we will use some of the language of massive gravity.
The disorder is made manifest through the addition 
in the field theory side, of 
a random-field term to the Hamiltonian: \begin{equation}
H_{\mathrm{rf}} = \int \mathrm{d}^d\mathbf{x} \; g(\mathbf{x}) \mathcal{O}(\mathbf{x}),
\end{equation}where $g(\mathbf{x})$ is a ($t$-independent) Gaussian random variable: 
\begin{subequations}\label{eq:6}\begin{align}
\mathbb{E}[g(\mathbf{k})]&=0, \\
\mathbb{E}[g(\mathbf{k})g(\mathbf{q})] &= \varepsilon^2 \delta(\mathbf{k}+\mathbf{q}).
\end{align}\end{subequations}
Here $\mathcal{O}$ is the operator dual to the scalar field $\psi$ introduced above and $\varepsilon$ is a small dimensionful number characterizing the scale of the disorder.  Note that this disorder will locally, but not globally, violate the $\mathbb{Z}_2$ symmetry $\psi \rightarrow -\psi$ (corresponding to $\mathcal{O}\rightarrow -\mathcal{O}$ in the field theory).   We choose $\mathcal{O}$ to be a relevant operator even with random field disorder.   As we derive shortly, this leads to a hyperscaling-violating generalization of the Harris criterion \cite{hartnollimpure}: \begin{equation}
{\Delta < \frac{d-\theta}{2}+z.}  \label{eqharris}
\end{equation} Disorder may thus be treated perturbatively in the UV; disorder is relevant in the IR, but we use  
finite temperature to serve as an IR regulator, allowing us to 
treat disorder perturbatively \emph{everywhere}.  We will discuss the IR as $T\rightarrow 0$ in more detail below.
Due to scattering off of the random 
field disorder, we expect that $\rho_{\mathrm{dc}} \sim \varepsilon^2$.   

\section{Holography}
We now discuss our holographic computations related to the computation of $\rho_{\mathrm{dc}}$.  We proceed in three steps:  first, we use holography to derive the Harris criterion, as advertised.    Then, we compute $\rho_{\mathrm{dc}}$ using the massive gravity analogy.  Finally, we discuss the breakdown of perturbation theory.

\subsection{The Harris Criterion and a Dirty Black Hole}

From standard holography, we immediately see that the Gaussian variable $ g(\mathbf{x})$ can be directly translated to the source of $\psi(\mathbf{x})$. 
  In order to compute $\rho_{\mathrm{dc}}$, we therefore perturbatively construct a statistical ensemble of EMD black holes with sourced scalar hair, one for each value of the source $\psi(\mathbf{x})$, and then take the statistical average.  From this black hole with ``dirty'' scalar hair, we then compute $\rho_{\mathrm{dc}}$ using the technique of \cite{blaketong1, blaketong2}.   Before beginning, we must ensure that the scalar hair is perturbative in both the UV and the IR, and so we must find a generalization of the Harris criterion to hyperscaling-violating theories.   This can be seen by an elegant holographic argument:   the contribution to the stress tensor $T_{MN}$ from the scalar fields must be small compared to $R_{MN}$ (e.g.) in the UV.   For a hyperscaling-violating geometry we have $R_{rr}\sim r^{-2}$; the contribution from the scalar fields will be $r^{2\nu_-- d^2/(d-\theta) - 2}$.  We conclude when \comment{$\nu_- > d^2/2(d-\theta)$} the disorder will be perturbative.   This results in the generalized Harris criterion, Eq. (\ref{eqharris}).

If the disordered hair is perturbative, to leading order in $\varepsilon$, we can simply solve Eq. (\ref{eqpsi0}) to determine the $\psi$ background. The correct solution is the one which is regular in the interior deep IR of the geometry.
{\em E.g.\/} at $T=0$ the solution is
\begin{align}
\label{eq:5}
\psi_0 &= g(\mathbf{k}) r^{\nu_-} + \cdots = \mathcal{C}r^{\frac{d}{2}(1+\frac{z}{d-\theta})} \mathrm{K}_{(1-\theta/d)(\nu_+-\nu_-)/2}\left(C_{d,\theta} |k| r^{d/(d-\theta)}\right)
\end{align}
$\mathcal{C}$ is dependent on $k$ and $g(\mathbf{k})$ and is chosen to ensure the correct UV scaling.   At finite $T$, the solution will be modified slightly, although this description is quantitatively accurate for large momentum modes.  To leading order in $\varepsilon$ this is a complete solution \cite{blaketong2}; corrections to EMD fields are $\sim\varepsilon^2$.   Although this is the same order as $\rho_{\mathrm{dc}}$, the inhomogeneous corrections cannot affect $\rho_{\mathrm{dc}}$, and the homogeneous corrections are subleading to the background, so for the purposes of computing $\rho_{\mathrm{dc}}$, we can treat the EMD background as unchanged \cite{blaketong2}.   

\subsection{DC conductivity}

The analytic computation of $\rho_{\mathrm{dc}}$ due to a scalar perturbation at a single fixed momentum $k_{\mathrm{L}}$ has been shown in \cite{blaketong2}.
We will generalize their formalism to an infinite number of random momentum modes with the distribution Eq. \eqref{eq:6}.
It is not entirely obvious that this generalization is possible. A calculation of the conductivity naively requires considering coupling a spatial component of the gauge field $\delta A_x$ to all spin 1 moments of the distribution $\int \mathrm{d}^d \mathbf{k} k_x k^{2n} \delta \psi$
.\footnote{Note that this infinite tower automatically collapses for a single momentum mode, as $\partial^2 \cos(kx) = -k^2 \cos(kx)$, so all of these modes are proportional.  This is not true when we have modes at different momentum.
}
We will see that a judicious choice of scalar perturbations effectively reduces the number of spin 1 perturbations to three as before. We will also find that we can compute $\rho_{\mathrm{dc}}$ before averaging over the disorder.

We 
proceed. The conductivity follows from the response to a finite frequency, zero-momentum perturbation $\delta A_x(\omega, \mathbf{k}=0,r)$. As in \cite{blaketong2} this perturbation couples to $\delta \tilde{g}_{tx}(\omega,r) = \delta g_{tx}r^2/L^2$, $\delta\tilde{g}_{rx}(\omega,r)$, and 
\begin{equation}
\delta \psi(\omega,\mathbf{k}\neq 0,r) = \psi_0(\mathbf{k},r) \delta P(\omega,\mathbf{k},r).
\end{equation}
where $\psi_0(\mathbf{k},r)$ is the perturbative solution in Eq. \eqref{eq:5}.
To lowest order none of these couple to dilaton perturbation, despite the nontrivial functions $Z(\Phi), V(\Phi), B(\Phi)$, because the dilaton is a spin zero mode, and the dilaton background is at zero spatial momentum.   

Following \cite{blaketong2} we can set $\delta\tilde{g}_{rx}=0$ by a gauge choice. Its corresponding equation of motion ---the $rx$-component of Einstein's equations--- is a constraint. Projecting on the zero-momentum mode one finds
\begin{align}
  \hat{\mathcal{Q}}\delta A_x - L\kappa e \delta\mathcal{P}_x = \frac{eL\delta \tilde{g}_{tx}^\prime}{r^d\kappa\sqrt{GH}}\label{eq22}
\end{align}
where we have defined \begin{equation}
\frac{f}{\omega r^d}\sqrt{\frac{H}{G}} \int \frac{\mathrm{d}^d\mathbf{k}}{(2\pi)^d}\; k_x\psi_0(\mathbf{k},r)^2 \delta P(\omega,\mathbf{k},r)^\prime \equiv \delta \mathcal{P}_x(\omega, r).  \label{scalareq1}
\end{equation}
In deriving Eq. (\ref{eq22}) the contribution
 proportional to $\psi_0\psi_0^\prime \delta P$ which survives if $\delta P$ is a constant has been ignored. It should be considered as an $\epsilon^2$ contribution to the background, whereas we only keep terms up to $\epsilon$.   
The other equations are the $x$ component of
Maxwell's equation: 
\begin{equation}
\left(\frac{eL}{\kappa} \delta \tilde{g}_{tx} \hat{\mathcal{Q}} - r^{2-d}\sqrt{\frac{H}{G}}fZ\delta A_x^\prime \right)^\prime + r^{2-d}Z\sqrt{\frac{G}{H}} \frac{\omega^2}{f} \delta A_x= 0, \label{eq:7}
\end{equation}
and the scalar equation 
\begin{equation}
-\sqrt{\frac{G}{H}}\frac{\psi_0(\mathbf{k})^2 k_x \omega \delta \tilde{g}_{tx}}{fr^d}
= \left(\frac{f\psi_0(\mathbf{k})^2 \delta P(\mathbf{k})^\prime}{r^d}\sqrt{\frac{H}{G}}\right)^\prime+ \frac{\omega^2}{f r^d}\sqrt{\frac{G}{H}} \psi_0(\mathbf{k})^2 \delta P(\mathbf{k})    \label{scalareq2}
\end{equation}
The $rt$-component of Einstein's equations is not independent and follows from the previous equations.

The key observation is as follows: ``averaging'' the scalar equation over its momentum distribution with weight $k_x$ $\int \frac{\mathrm{d}^d \mathbf{k} }{(2\pi)^d} k_x$, we can turn it into 
\begin{equation}
\delta \mathcal{P}_x^\prime = -\frac{ \delta\tilde{g}_{tx}}{dfr^d} \sqrt{\frac{G}{H}} \int \frac{\mathrm{d}^d\mathbf{k}}{(2\pi)^d} \; k^2\psi_0(\mathbf{k})^2  - \frac{\omega}{ f r^d}\sqrt{\frac{G}{H}}\int \frac{\mathrm{d}^d\mathbf{k}}{(2\pi)^d}\; k_x \psi_0(\mathbf{k})^2 \delta P(\mathbf{k})\label{eq:9}
\end{equation}
In the first term on the right-hand side we have used isotropy of the random disorder to substitute $k^2/d$ for $k_x^2$.

For the dc-conductivity we wish to know the $\omega \rightarrow 0$ solution to these equations. This limit is subtle, due to the presence of the horizon where $f(r_h)=0$. Note, however, that away from the horizon, where in the $\omega \rightarrow 0$ limit we can ignore the higher order $\omega$ contributions in \eqref{eq:7} and \eqref{eq:9},
the system of equations closes to a finite set of differential equations for $\delta A_x$, $\delta\mathcal{P}_x$, and $\delta \tilde{g}_{tx}$. 

We now proceed to compute the conductivity following the steps in  \cite{blaketong1}, generalized to higher dimensions.   
Integrating once, Eq. \eqref{eq:7} is equal to
\begin{align}
  \label{eq:8}
  \left( r^{2-d}\sqrt{\frac{H}{G}}fZ\delta A_x^\prime  -  \delta \tilde{g}_{tx} \frac{eL\hat{\cQ}}{\kappa}\right) = C -\int\mathrm{d}r \; r^{2-d}Z\sqrt{\frac{G}{H}} \frac{\omega^2}{f} \delta A_x 
\end{align}
in terms of an unknown integration constant $C$. We eliminate $\delta\tilde{g}_{tx}$ using the scalar equation of motion Eq. \eqref{eq:9} and obtain 
\begin{align}
C &=\sqrt{\frac{H}{G}} f \left[ r^{2-d}Z\delta A_x^\prime  + {\delta \mathcal{P}_x^\prime} \frac{eL\hat{\cQ}}{\kappa} dr^d \left(\int \frac{\mathrm{d}^d\mathbf{k}}{(2\pi)^d}\; k^2 \psi_0(\mathbf{k},r)^2\right)^{-1}\right] \nonumber\\
&~~ +\omega^2\sqrt{\frac{G}{H}} \frac{1}{fr ^d}\left[\int^r r^{2}Z \delta A_x +\frac{eL\hat{\cQ}}{\omega \kappa}  \int \frac{\mathrm{d}^d \mathbf{k}}{(2\pi)^d} \psi_0(\mathbf{k},r)^2 \delta P(\mathbf{k}) \right].
\label{eq25}
\end{align} 
We now show that the constant $C(\omega, \mathbf{k})$ is proportional to the dc-conductivity.
Note that in the derivation of Eq. \eqref{eq:8} and \eqref{eq25} we have only used the form of the metric and the background solution, but not any specific expressions. In particular, a full solution interpreting from an asymptotically AdS boundary to an hyperscaling violating quantum critical IR will have solution that is of exactly the same form. For the background we now take such a fully asymptotically AdS completed solution, and evaluate the solution near the AdS-boundary. There $f\approx G\approx H\approx Z\approx 1$ as $r\rightarrow 0$. Consider first Eq. \eqref{eq22}. As $\psi_0$ by construction corresponds to a relevant operator,  $\psi_0$ behaves as $\psi_0= g(\mathbf{k}) r^{\Delta_{\mathrm{UV}}}+\ldots$ with $\Delta_{\mathrm{UV}} >0$, it follows that $\delta\tilde{g}_{tx} \sim r^{d+1}$, as $\delta A_x \sim r^{0}$.\footnote{Note that this is precisely the expected scaling for $\delta \tilde{g}_{tx}$ in the absence of a source.} Consider then Eq. \eqref{eq:8}. It means that $\delta\tilde{g}_{tx}$ is always subleading near $r\rightarrow 0$ and we can solve for the AdS-boundary behavior of the fluctuation $\delta A_x =  C_0+\frac{1}{d-1}C r^{d-1}+\ldots$. The AdS/CFT dictionary tells us that
the dc-conductivity is equal to
\begin{align}
  \label{eq:12}
  \sigma_{\mathrm{dc}} = \frac{1}{e^2} \lim_{\omega \rightarrow 0} \frac{-1}{\mathrm{i}\omega} \lim_{r\rightarrow 0}r^{2-d}\frac{ \delta A_x^\prime(r)}{\delta A_x(r)} = \frac{1}{e^2} \lim_{\omega \rightarrow 0} \frac{-1}{\mathrm{i}\omega} \frac{C}{C_0},
\end{align}
and therefore
\begin{equation}
C=-\mathrm{i}\omega \sigma_{\mathrm{dc}} e^2 \delta A_x(r=0).
\end{equation}

The coefficient $C$ can be evaluated at the horizon, as follows.   We know that, near the horizon, where $f(r) \sqrt{H/G} \sim T(r-r_{\mathrm{h}})+\ldots$ : \begin{equation}
\label{eq:16}
\delta A_x,~ \delta\mathcal{P}_x,~\frac{\delta\tilde{g}_{tx}}{f(r)}\sim \left(f\sqrt{\frac{H}{G}}\right)^{-\mathrm{i}\omega/4\pi T} \sim \left(T(r_{\mathrm{h}}-r)\right)^{-\mathrm{i}\omega/4\pi T}.  
\end{equation}
For $\delta A_x \sim \mathrm{O}(1)$, it then follows that near the horizon $\delta A_x',~\delta\tilde{g}_{tx},~\delta\mathcal{P}_x'\sim \omega$. Therefore, to leading order in $\omega$, as $\omega\rightarrow 0$, only the first line of Eq. (\ref{eq25}) contributes. Now taking the limit $\omega \rightarrow 0$ the near-horizon limit of Eq. \eqref{eq22}  reduces to 
\begin{align}
  \label{eq:13}
  \hat{\mathcal{Q}} \delta A_x(r=r_{\mathrm{h}}, \omega=0) = Le\kappa \; \delta \mathcal{P}_x (r=r_{\mathrm{h}},\omega=0) .
\end{align}
Thus
\begin{align}
  \label{eq:14}
  C = \lim_{r\rightarrow r_{\mathrm{h}}} \sqrt{\frac{H}{G}} f \left[  r^{2-d} Z + \frac{\hat{\cQ}^2}{\kappa^2}dr^d \left(\int \frac{\mathrm{d}^d\mathbf{k}}{(2\pi)^d}\; k^2 \psi_0(\mathbf{k},r)^2\right)^{-1} \right] \delta A_x^\prime(r=r_{\mathrm{h}},\omega=0).
\end{align}
Substituting for the asymptotic behavior of $\delta A_x^\prime$ near the horizon given in Eq. \eqref{eq:16}, we find 
\begin{align}
  \label{eq:17}
  C = -\mathrm{i} \omega \left[  r_{\mathrm{h}}^{2-d} Z + \frac{\hat{\cQ}^2}{\kappa^2}dr_{\mathrm{h}}^d \left(\int \frac{\mathrm{d}^d\mathbf{k}}{(2\pi)^d}\; k^2 \psi(\mathbf{k},r_{\mathrm{h}})^2\right)^{-1} \right] \delta A_x (r_h).
\end{align}
%
%
%
%
%
%
%
We conclude that, as $\varepsilon$ is small, to leading order in $\varepsilon$: \begin{equation}
\rho_{\mathrm{dc}} = \frac{1}{\sigma_{\mathrm{dc}}} \sim \left(\int \mathrm{d}^d\mathbf{k}\; k^2 \psi(k,r_{\mathrm{h}})^2\right) r_{\mathrm{h}}^{-d} \frac{\delta A_x(r=0,\omega=0)}{\delta A_x(r=r_{\mathrm{h}},\omega=0)}.
\end{equation}Now, the fact that $C \sim \omega$ implies that, $\delta A_x^\prime \sim \omega$, or that $\delta A_x$ is, to leading order in $\omega$, independent of $r$.\footnote{This assumes that, generically in the bulk, $\delta A_x^\prime$ and $\delta \mathcal{P}_x^\prime$ do not cancel each other.   See \cite{blaketong1, blaketong2} for more.}   Noting that $s\sim r_{\mathrm{h}}^{-d}$, we obtain \begin{equation}
\rho_{\mathrm{dc}} \sim s \int \mathrm{d}^d\mathbf{k}\; k^2 \psi(k,r_{\mathrm{h}})^2 \equiv s\, m^2(r_{\mathrm{h}}).
\end{equation}
In analogy with \cite{blaketong2}, we have noted this is an effective graviton mass.   

Note that because $\rho_{\mathrm{dc}} \sim m^2 \sim \varepsilon^2$, as $\varepsilon \rightarrow 0$, the resistivity vanishes.   This is consistent with the fact that  $\mathrm{Re}(\sigma(\omega)) \sim \delta(\omega) + \cdots$ at small $\omega$ when there is translational symmetry and finite charge density.  Holography can be used to compute $\sigma_{\mathrm{dc}}(\omega)$ at finite $\omega$ as well -- however, the massive gravity analogy will require modifications:  we can see from Eq. (\ref{scalareq2}) that the scalar equation does not close to an equation for $\delta \mathcal{P}_x$ unless $\omega\rightarrow 0$.

The remaining task is to determine $m^2(r_{\mathrm{h}})$. 
To do so we need to evaluate $\psi_0(\mathbf{k},r)$ at the horizon $r=r_h$. For momenta where $k \gg r_h^{-d/(d-\theta)}$, i.e. $k \gg T^{1/z}\mathcal{Q}^{1/(d-\theta)}$, we may neglect the effect of temperature and approximate $\psi_0(\mathbf{k},r)$ with its $T=0$ Bessel function solution Eq. \eqref{eq:5}. For these momenta the Bessel function is exponentially small at $r=r_{\mathrm{h}}$, and we can ignore their contribution. $T^{1/z}$ thus serves as an effective UV cut-off in the momentum integral in $m^2$.   The integral over $k$ will give us an overall factor of $T^{(d+2)/z}$ -- as we will see, the scaling due to $\psi^2$ is approximately independent of $k$ in this regime.

For the remaining modes $k \ll T^{1/z}\mathcal{Q}^{1/(d-\theta)}$ we evaluate $\psi_0(\mathbf{k},r)$ by a matching procedure. For these solutions the presence of the horizon is relevant. Near the horizon, the equation of motion for the background $\psi_0$ becomes
\begin{align}
  \label{eq:10}
  &\partial_r^2 \psi + \frac{1}{r-r_{\mathrm{h}}} \partial_r \psi - \frac{M^2}{(1-r/r_{\mathrm{h}})}=0~,\\
~~ &M^2 \equiv \frac{k^2 G(r_{\mathrm{h}}) + B_0/ r_{\mathrm{h}}^2}{d(1+z/(d-\theta))}.
\end{align}
The solution regular at the horizon is the Bessel function
\begin{align}
  \label{eq:11}
  \psi_{\mathrm{near-hor}}(\mathbf{k},r) = \beta \mathrm{I}_0 \left(2M r_{\mathrm{h}}\sqrt{1-\frac{r}{r_{\mathrm{h}}}} \right).
\end{align}
For the small momenta range of interest  $k \lesssim r_{\mathrm{h}}^{-d/(d-\theta)}$, $M r_{\mathrm{h}}$ is essentially a number independent of temperature. Therefore at a matching point $r \sim r_{\mathrm{h}}$, the Bessel function has no non-trivial scaling. Knowing that $\psi_{\mathrm{far}} \sim r^{\nu_-}$, we determine $\beta \sim r_{\mathrm{h}}^{\nu_-}$.    Note that this estimate is independent of $k\lesssim T^{1/z}$.
 It is straightforward from here to recover the full temperature dependence of the graviton mass: \begin{equation}
m^2 \sim \frac{\rho_{\mathrm{dc}}}{s} \sim T^{(d+2 - 2(1-\theta/d)\nu_-)/z} \sim T^{(2-d + 2\Delta - 2z+\theta)/z}.
\end{equation}Evidently, $\rho_{\mathrm{dc}}/s$ generically carries temperature dependence, showing that a conjecture of \cite{dsz} only holds in special cases.  Studying $\rho_{\mathrm{dc}}$ directly, we find 
\begin{equation}
\rho_{\mathrm{dc}}\sim \varepsilon^2 T^{2(1+\Delta-z)/z} + \mathrm{O}\left(\varepsilon^4\right),
\end{equation} 
as we quoted in Eq. (\ref{rhodc}).

It is useful to express our main result in Eq.~(\ref{rhodc}) in a condensed matter notation. It is conventional to determine
the scaling dimension, $\Delta$, of the ``order parameter'' $\mathcal{O}$ coupling to the random field by its ``anomalous'' dimension $\eta$.
For a theory with dynamic scaling exponent $z$, the relationship between $\Delta$ and $\eta$ is \cite{sachdevbook}
\begin{align}
  \label{eq:11a}
  \Delta = \frac{d+z-2+\eta}{2}.
\end{align}
Then Eq.~(\ref{rhodc}) becomes
\begin{equation}
\rho_{\mathrm{dc}} \sim \varepsilon^2 T^{(d-z+\eta)/z},  \label{rhodca}
\end{equation}
a result quoted in Ref.~\cite{raghu}.

\subsection{Breakdown of Perturbation Theory}

It is also worth asking for what value of $\varepsilon$ we expect perturbation theory to break down.   To do this, we check when the scalar hair non-perturbatively back-reacts on the geometry: i.e., when is the $\psi$ contribution to Einstein's equations of the same order as the contributions of the solution we are perturbing around.   A quick check at $r\sim r_{\mathrm{h}}$ reveals that all components of Einstein's equations break down at the same scale, if the disorder strength is strong enough.   For example, using the $xx$ component of Einstein's equations at $r\sim r_{\mathrm{h}}$, the scalar backreaction becomes nonperturbative when\begin{equation}
R_{xx} \sim r_{\mathrm{h}}^{-2d/(d-\theta)} \sim T^{2/z} \sim \int \mathrm{d}^dk\; k^2 \psi^2 \sim m^2,
\end{equation}or when the temperature falls below 
\begin{equation}
T^{(z-\Delta+(d-\theta)/2)/z} \lesssim \varepsilon   \label{eq49}
\end{equation}
Because the dilaton couples in a universal, exponential manner to each term in the matter stress tensor in the IR, the dilaton equation of motion will break at the same scale.

It is easy to check, given this result, that it is impossible to have a regime where we can trust the calculation where $\rho_{\mathrm{dc}} \rightarrow \infty$ (i.e., the strange metal becomes an insulator) as $T\rightarrow 0$, without the perturbative approximation breaking down. When perturbation theory breaks down, at $\varepsilon \sim T^{(z-\Delta+(d-\theta)/2)/z}$  we universally find 
\begin{equation}
\rho_{\mathrm{dc}} \sim T^{(2+d-\theta)/z} \sim T^{2/z}s,   
\label{rhodc2}
\end{equation} 
independent of the choice of $\Delta$.
This is in fact the scaling one finds for $\varepsilon$ fixed and $\Delta= \frac{d-\theta}{2}+z$ saturating the Harris bound. It self-consistently shows that the effect of random-field disorder from operators with dimensions that violate the Harris criterion is always non-perturbative.
Comparing to the conjecture of \cite{dsz} we find agreement in the limit $z\rightarrow \infty$, up to a possible logarithmic correction, despite the fact that Eq. (\ref{rhodc}) appears to badly violate $\rho_{\mathrm{dc}}\sim s$.

Interestingly, this result also qualitatively agrees with a memory matrix based argument for the $\mathrm{AdS}_4$-Reissner-N\"ordstrom geometry ($z=\infty$, $\theta=0$), which found that $\rho_{\mathrm{dc}} \sim (\log T^{-1})^{-1}$ due to random-field disorder \cite{hartnollhofman}.\footnote{This agreement is especially interesting, as \cite{hartnollhofman} used irrelevant operators to add disorder, whereas we used relevant operators.}   
Due to the presence of $1/z$ corrections, it is natural to expect such logarithmic correction factors to appear in Eq. (\ref{rhodc2}) as well.

One might ask why we could ignore the first term in Eq. (\ref{eq:17}) in this argument.   It is straightforward to check that the first term is comparable to the second term precisely at the same scale as perturbation theory breaks down; thus, in the regime of validity of the perturbative massive gravity analogy, we can ignore this contribution to $\rho_{\mathrm{dc}}$.

\section{Memory Matrix Method}

We will now confirm our holographic computation with an independent calculation via the memory matrix method \cite{forster}, which is especially suited to the computation of transport quantities in the absence of long-lived quasiparticles \cite{hkms,jung,hartnollhofman}. 
The basic procedure was reviewed recently in \cite{raghu}, and the main result for the resistivity is
\begin{equation}
\rho_{\mathrm{dc}} \sim \varepsilon^2 \int\limits_0^{T^{1/z}} \mathrm{d}^dk\; k^2 \lim_{\omega\rightarrow 0} \mathrm{Im}\frac{G^{\mathrm{R}}_{\mathcal{O}\mathcal{O}}(\omega,k)}{\omega}.   \label{eq50}
\end{equation}The integral over $k$, and the $k^2$ factor, give $T^{(d+2)/z}$.   Using the fact that for $k\sim T^{1/z}$ \cite{sachdevye, sachdev2} \begin{equation}
\lim_{\omega\rightarrow 0} \mathrm{Im}\frac{G^{\mathrm{R}}_{\mathcal{O}\mathcal{O}}(\omega,k)}{\omega} \sim T^{(2\Delta-2z-d)/z},
\end{equation}we arrive at Eq. (\ref{rhodc}).  


\section{Conclusions}
In conjunction with the recent result \cite{raghu}, our findings show that random-field disorder can have an extremely strong effect on the low-temperature dc-conductivity.  Unless there is a mechanism which protects transport from random-field scattering, at low-temperatures random-field disorder must always be taken into account.    In particular, regardless of disorder strength, at low enough temperatures disorder due to relevant operators leads to non-perturbative effects in the IR \cite{robertson, adrian, erica, laimei}.

We noted that at the breakdown of perturbation theory $\rho_{\mathrm{dc}}\sim T^{2/z} s$. For $z=\infty$ this reduces to a linear relation between the dc-resistivity and the entropy density. It would be interesting if there is a deep reason why this must be the case.

The qualitative agreement in $T$-scaling between the effective graviton mass calculation, and the memory matrix formalism, has been shown for a single momentum mode in \cite{blaketong2}.   
Quantitatively we have shown that the agreement between the effective graviton mass calculation, and the memory matrix formalism remains 
for a generic scaling theory with finite values of $z$ and $\theta$ and for disorder, and when the memory matrix calculation is completely independent of holography.
Note that the agreement of these two calculations is \emph{not} a trivial consequence of dimensional analysis -- $\mathcal{Q}/T^{d/z}$ is a dimensionless quantity.    

Although the memory matrix method appeared substantially faster, the holographic method contains its own advantages. In particular, we are able to determine the disorder strength at which perturbation theory breaks down.   Holographic methods also allow, in principle, a determination of results to all orders in the disorder strength \cite{hartnollsantos}.  

Looking forward, it would be interesting to extend these results to a quantum field theory which is manifestly UV-completed to a conformal field theory, or by duality, studying a geometry which is UV-completed to AdS.  As \cite{raghu} recently noted, such a UV-completion may provide another universal mechanism for $\rho_{\mathrm{dc}}\sim T$ at high temperatures, without the requirement of local criticality.   In addition, 
  it would be interesting to 
  determine the optical (finite frequency) resistivity due to random-field disorder.

\section*{Acknowledgements.}
We thank Richard Davison, Blaise Gout\'eraux and the anonymous reviewer for discussions.
The research was supported by the U.S.\ National Science Foundation under grant DMR-1103860 and by the Templeton Foundation.
 A.L. is supported by the Smith Family Graduate Science and Engineering Fellowship at Harvard University. K.S. is supported in part by a VICI grant of the
Netherlands Organization for Scientific Research (NWO), by the
Netherlands Organization for Scientific Reseach/Ministry of Science
and Education (NWO/OCW) and by
the Foundation for Research into Fundamental Matter (FOM).

\end{document}